\documentclass{article}

\usepackage{PRIMEarxiv}

\usepackage[utf8]{inputenc} 
\usepackage[T1]{fontenc}    
\usepackage{hyperref}       
\usepackage{url}            
\usepackage{booktabs}       
\usepackage{amsfonts}       
\usepackage{nicefrac}       
\usepackage{microtype}      
\usepackage{lipsum}
\usepackage{fancyhdr}       
\usepackage{graphicx}       
\graphicspath{{media/}}     
\usepackage{todonotes}

\usepackage{amsmath}

\newcommand{\norm}[1]{\left\lVert#1\right\rVert}
\newcommand{\diag}[1]{\mathrm{diag}\left(#1\right)}
\newcommand{\DKL}[2]{D_{\mathrm{KL}}\left(#1 \parallel #2 \right)}
\newcommand{\DJe}[2]{D_{\mathrm{J}}\left(#1 \parallel #2 \right)}

\newcommand{\ab}{\boldsymbol{a}}
\newcommand{\bb}{\boldsymbol{b}}
\newcommand{\cb}{\boldsymbol{c}}
\newcommand{\xb}{\boldsymbol{x}}
\newcommand{\zb}{\boldsymbol{z}}
\newcommand{\gb}{\boldsymbol{g}}
\newcommand{\hb}{\boldsymbol{h}}
\newcommand{\vb}{\boldsymbol{v}}

\newcommand{\mb}{\boldsymbol{m}}

\newcommand{\sbo}{\boldsymbol{s}}

\newcommand{\epsb}{\boldsymbol{\epsilon}}

\newcommand{\lambdab}{\boldsymbol{\lambda}}
\newcommand{\thetab}{\boldsymbol{\theta}}
\newcommand{\mub}{\boldsymbol{\mu}}
\newcommand{\sigmab}{\boldsymbol{\sigma}}
\newcommand{\xib}{\boldsymbol{\xi}}
\newcommand{\rhob}{\boldsymbol{\rho}}

\newenvironment{eq}{\begin{equation}\begin{aligned}}{\end{aligned}\end{equation}\ignorespacesafterend}

\makeatletter
\newcommand\thefontsize[1]{{#1 The current font size is: \f@size pt\par}}
\makeatother

\usepackage{changepage}
\usepackage{enumitem}
\usepackage{xcolor}

\usepackage[style=numeric, sorting=none, giveninits=true, isbn=false,eprint=false, minnames=15, maxnames=20, backend=biber]{biblatex}
\bibliography{main}  
\AtEveryBibitem{\clearfield{note}}
\AtEveryBibitem{\clearlist{language}}
\AtEveryBibitem{\clearfield{urlyear}}
\AtEveryBibitem{\ifentrytype{misc}{%
  }{%
    \clearfield{url}%
  }%
}

\definecolor{cblue}{rgb}{0.050383, 0.029803, 0.527975}
\definecolor{cred}{rgb}{0.881443, 0.392529, 0.383229}
\definecolor{cyellow}{rgb}{0.988648, 0.809579, 0.145357}

\hypersetup{
     linkbordercolor=cblue,
     citebordercolor=cred,   
     urlbordercolor=cyellow
}

\usepackage{algorithm}
\usepackage{algpseudocode}
\algnewcommand\And{\textbf{and}}
\algrenewcommand\algorithmicrequire{\textbf{Initialization:}}

\usepackage{siunitx}
\title{Dynamic Learning Rate Decay for Stochastic Variational Inference}

\author{
  Maximilian Dinkel\\
  Institute for Computational Mechanics \\
  Technical University of Munich \\
  \texttt{\{maximilian.dinkel\}@tum.de} \\
  \And
  Gil Robalo Rei \\
  Institute for Computational Mechanics\\
  Technical University of Munich \\
  \texttt{\{gil.rei\}@tum.de} \\
  \And
  Wolfgang A. Wall \\
  Institute for Computational Mechanics, Munich Data Science Institute\\
  Technical University of Munich \\
  \texttt{\{wolfgang.a.wall\}@tum.de} \\
}

\begin{document}
\maketitle
\delimitershortfall=-1pt
\setlength{\parindent}{0pt}

\begin{abstract}
Like many optimization algorithms, Stochastic Variational Inference (SVI) is sensitive to the choice of the learning rate. 
If the learning rate is too small, the optimization process may be slow, and the algorithm might get stuck in local optima. On the other hand, if the learning rate is too large, the algorithm may oscillate or diverge, failing to converge to a solution. Adaptive learning rate methods such as Adam, AdaMax, Adagrad, or RMSprop automatically adjust the learning rate based on the history of gradients.
Nevertheless, if the base learning rate is too large, the variational parameters might still oscillate around the optimal solution. With learning rate schedules, the learning rate can be reduced gradually to mitigate this problem.
However, the amount at which the learning rate should be decreased in each iteration is not known a priori, which can significantly impact the performance of the optimization. \\ 
In this work, we propose a method to decay the learning rate based on the history of the variational parameters. We use an empirical measure to quantify the amount of oscillations against the progress of the variational parameters to adapt the learning rate. The approach requires little memory and is computationally efficient. We demonstrate in various numerical examples that our method reduces the sensitivity of the optimization performance to the learning rate and that it can also be used in combination with other adaptive learning rate methods.
\end{abstract}

\keywords{Variational Inference, Learning Rate}

\section{Introduction}

\begin{figure}[h!]
    \centering
    \includegraphics[]{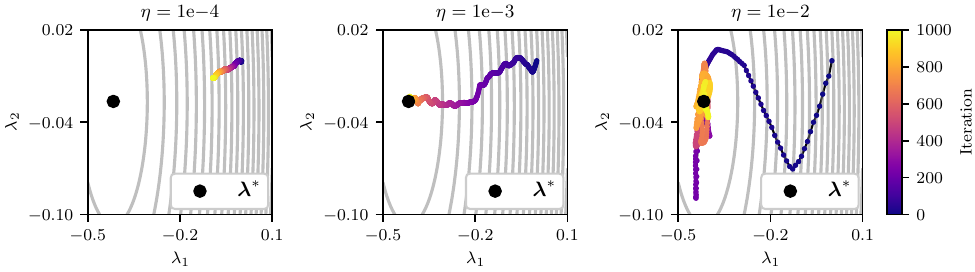}
    \caption{Optimization progress of SVI for a toy problem with two variational parameters $\lambdab$ using Adam with different learning rates $\eta$.}
    \label{fig:1d_convergence}
\end{figure}

Stochastic Variational Inference (SVI) has emerged as a powerful tool for approximating complex posterior distributions in Bayesian inference across various applications. By casting inference as an optimization problem, SVI enables scalable and efficient posterior approximation, particularly in high-dimensional probabilistic models \cite{hoffman_stochastic_2013, blei_variational_2017}. Crucial to the optimization process in SVI is the choice of the learning rate, a hyperparameter that governs the step size taken in the parameter space during optimization. An inadequate learning rate can severely impede the performance of SVI \cite{ranganath_adaptive_2013}. If set too low, the optimization progresses slowly, and the algorithm may become trapped in local optima. Conversely, a high learning rate can cause the algorithm to diverge or oscillate around the optimum. The oscillations around the optimum occur due to the noise of the gradient estimate, and the variance of these oscillations is proportional to the learning rate \cite{lecun_efficient_1998}. To address this issue, adaptive learning rate methods like Adam, AdaMax \cite{kingma_adam_2014}, Adagrad \cite{duchi_adaptive_2011}, and RMSprop \cite{tieleman_lecture_2012} automatically adjust the learning rate based on the gradients' history.
While these methods can enhance the optimization process, they are not immune to problems arising from a high base learning rate, which can still lead to oscillations around the optimal solution. 
Additionally, first-order optimization methods rely solely on gradient information, which might not fully capture the curvature of the parameter space, potentially slowing down convergence. Second-order optimization methods, such as natural gradient descent, offer a promising alternative by incorporating curvature information to achieve more efficient and stable convergence. Natural gradient descent, in particular, adjusts the update direction based on the Fisher information matrix, which may lead to faster convergence \cite{amari_natural_1998}. Despite its advantages, the computation of the Fisher information matrix can be computationally expensive, and the sensitivity to the learning rate still exists.
\\ Figure \ref{fig:1d_convergence} shows the optimization progress of SVI for a toy problem with only two variational parameters using Adam. It can be observed that for a large base learning rate of $\eta=\num{1e-2}$ the optimizer reaches the vicinity of the optimal solution quickly but fails to converge due to oscillations. With a lower learning rate of $\eta=\num{1e-3}$ the oscillating behavior around the optimal solution is significantly reduced. However, it also takes significantly more iterations to reach the vicinity of the optimal solution. With an even lower learning rate of $\eta=\num{1e-4}$, the algorithm does not even reach the vicinity of the optimal solution within the considered amount of iterations. \\
Learning rate schedules provide an alternative or complement to adaptive learning rates by gradually reducing the learning rate over time. 
Yet, the optimal schedule for decreasing the learning rate is generally unknown a priori and can significantly influence the optimization progress. 
In this paper, we introduce a novel approach to adaptively decay the learning rate based on the history of the variational parameters. Our method quantifies oscillations relative to the progress of the variational parameters, enabling dynamic adjustments to the learning rate. This approach is designed to be both memory-efficient and computationally lightweight, making it suitable for a large variety of applications. \\
We validate our method through a series of numerical experiments, demonstrating its ability to mitigate the sensitivity of SVI to the learning rate. Our results indicate that the proposed technique not only enhances SVI with standard stochastic gradient descent but also complements existing adaptive learning rate methods. \\
The rest of the paper is structured as follows: In section \ref{sec:methodology}, we revise SVI and introduce our novel approach for dynamic learning rate decay. Subsequently, we test our approach on three numerical examples in section \ref{sec:examples}. Finally, a conclusion is presented in section \ref{sec:conclusion}.
\footnote{Generally, we use plain letters for scalar, boldface letters for vector-valued, and capital letters for matrix-valued quantities. The symbol $\odot$ denotes the elementwise multiplication, $\oslash$ the elementwise division, and $^{\circ}$ denotes the elementwise power. The code for this paper is implemented in our (soon) open-source code QUEENS \cite{noauthor_queensgeneral_2024}.}

\section{Methodology}
\label{sec:methodology}

SVI combines Variational Inference (VI) with stochastic optimization. In this section, we briefly recap the methodology of SVI, including the reparameterization trick. Subsequently, we present our novel approach for dynamic learning rate decay for SVI.

\subsection{Stochastic Variational Inference}

Variational Inference aims to approximate the true posterior distribution \( p(\zb \mid \xb) \) with a simpler, tractable distribution \( q_{\lambdab}(\zb) \), parameterized by \(\lambdab\). The goal is to find the variational parameters \(\lambdab\) that make \( q_{\lambdab}(\zb) \) as close as possible to the true posterior distribution. Typically, the closeness is measured using the reversed Kullback-Leibler (KL) divergence:
\begin{eq}
    \DKL{q_{\lambdab}(\zb)}{p(\zb \mid \xb)} 
    &= \int q_{\lambdab}(\zb) \left[ \log q_{\lambdab}(\zb) - \log p(\zb \mid \xb) \right]  d\zb \\
    &= \int q_{\lambdab}(\zb) \left[ \log q_{\lambdab}(\zb) - \log p(\zb, \xb) \right]  d\zb + \log p(\xb).
    \label{eqn:kl}
\end{eq}
Since $\log p(\xb)$ is independent of $\lambdab$, maximizing the evidence lower bound (ELBO), defined as:
\begin{eq}
    \mathcal{L}(\lambdab) = \mathbb{E}_{q_{\lambdab}(\zb)} \left[ \log p(\zb, \xb) - \log q_{\lambdab}(\zb) \right],
    \label{eqn:elbo}
\end{eq}
is equivalent to minimizing the KL divergence $\DKL{q_{\lambdab}(\zb)}{p(\zb \mid \xb)}$ \cite{bishop_pattern_2006}. \\
In most applications, the ELBO and its gradient with respect to the variational parameters is intractable. Nevertheless, the optimal variational parameters can be found using stochastic optimization \cite{hoffman_stochastic_2013, robbins_stochastic_1951}. We can obtain a noisy but unbiased estimator of the ELBO gradient $\nabla_{\lambdab}^{\mathrm{MC}} \mathcal{L}(\lambdab)$  using Monte Carlo samples \cite{ranganath_black_2014, mohamed_monte_2020}:
\begin{eq}
    \nabla_{\lambdab} \mathcal{L}(\lambdab) &= \mathbb{E}_{q_{\lambdab}(\zb)} \left[ \nabla_{\lambdab} \log q_{\lambdab}(\zb) \left( \log p(\zb, \xb) - \log q_{\lambdab}(\zb) \right) \right]  \\
    &\approx \frac{1}{S} \sum_{s=1}^{S} \nabla_{\lambdab} \log q_{\lambdab}\left(\zb_s\right) \left( \log p\left(\zb_s, \xb \right) - \log q_{\lambdab}\left(\zb_s\right) \right), \quad \zb_s \sim q_{\lambdab}(\zb),
    \label{eqn:grad_elbo_bb}
\end{eq}
where the number of samples $S$ is often referred to as the batch size. With stochastic gradient ascent, the variational parameters can be updated iteratively using the above gradient estimate of the ELBO:
\begin{eq}
        \lambdab_{i+1} = \lambdab_i + \eta \nabla_{\lambdab}^{\mathrm{MC}} \mathcal{L}(\lambdab_i),
\end{eq}
where $\eta$ denotes the learning rate. The gradient estimate in \eqref{eqn:grad_elbo_bb} exhibits a high variance, which leads to slow convergence, especially in high dimensions \cite{ranganath_black_2014}. The variance can be reduced using Rao-Blackwellization \cite{casella_rao-blackwellisation_1996} or control variates \cite{ross_simulation_2006, paisley_variational_2012}. \\
In this work, we employ the reparameterization trick \cite{kingma_auto-encoding_2022, kingma_variational_2015} to obtain low-variance gradient estimates of the ELBO. Instead of sampling \(\zb\) directly from \(q_{\lambdab}(\zb)\), we express \(\zb\) as a deterministic function of \(\lambdab\) and an auxiliary variable \(\epsb\) with a simple distribution that does not depend on \(\lambdab\).
For instance, if \( q_{\lambdab}(\zb) \) is a Gaussian distribution \(\mathcal{N}(\zb \mid \boldsymbol{\mu}, \Sigma)\), we can reparameterize \(\zb\) as:
\begin{eq}
    \zb(\epsb, \lambdab) = \boldsymbol{\mu} + L \epsb, \quad \epsb \sim \mathcal{N}(\mathbf{0}, I),
\end{eq}
where $L$ is a lower triangular matrix, obtained by the Cholesky decomposition of $\Sigma=LL^{\top}$.
This reparameterization allows us to rewrite the expectation in the ELBO as:
\begin{eq}
    \mathcal{L}(\lambdab) = \mathbb{E}_{p(\epsb)} \left[ \log p(\zb(\epsb, \lambdab), \xb) - \log q_{\lambdab}(\zb(\epsb, \lambdab)) \right].
    \label{eqn:elbo_rpvi}
\end{eq}
Consequently, a gradient estimate can be obtained as:
\begin{eq}
    \nabla_{\lambdab} \mathcal{L}(\lambdab) &= \mathbb{E}_{p(\epsb)} \left[ \nabla_{\zb} \left(\log p(\zb(\epsb, \lambdab), \xb) - \log q_{\lambdab}(\zb(\epsb, \lambdab)) \right) \nabla_{\lambdab} \zb(\epsb, \lambdab)  - \nabla_{\lambdab} \log q_{\lambdab}(\zb(\epsb, \lambdab)) \right]  \\
    &= \mathbb{E}_{p(\epsb)} \left[ \nabla_{\zb} \left(\log p(\zb(\epsb, \lambdab), \xb) - \log q_{\lambdab}(\zb(\epsb, \lambdab)) \right) \nabla_{\lambdab} \zb(\epsb, \lambdab) \right]  \\
    &\approx \frac{1}{S} \sum_{s=1}^{S} \nabla_{\zb} \left(\log p(\zb(\epsb_s, \lambdab), \xb) - \log q_{\lambdab}(\zb(\epsb_s, \lambdab)) \right) \nabla_{\lambdab} \zb(\epsb_s, \lambdab), \quad \epsb_s \sim p(\epsb), 
    \label{eqn:grad_elbo_rpvi}
\end{eq}
where we removed the high-variance score function term $\nabla_{\lambdab} \log q_{\lambdab}(\zb(\epsb, \lambdab))$, as its expectation w.r.t. $p(\epsb)$ is zero \cite{roeder_sticking_2017}.

\subsection{Stochastic Optimization}

Next to vanilla Stochastic Gradient Ascent (or Descent), many other stochastic optimization methods, such as AdaGrad, RMSProp, and Adam, have evolved to improve optimization efficiency. Essentially, each of these algorithms introduces a mechanism  based on the history of gradients. \\
Adagrad (Adaptive Gradient Algorithm) \cite{duchi_adaptive_2011} scales down the learning rate of each parameter individually based on the sum of the squares of previous gradients. The update step can be written as:
\begin{eq}
    \hb_i &= \sum_{j=1}^i \gb_j^{\circ 2}   \\
    \lambdab_{i+1} &= \lambdab_i + \eta \gb_i \oslash \sqrt{\hb_i + \varepsilon} , 
\end{eq}
where $\hb_i$ is the sum of the squares of the gradients up to iteration $i$, $\gb_i$ is the current gradient, and \( \varepsilon \) is a small constant to prevent division by zero.
A downside of this method is that the continuous accumulation of the squared gradients in the denominator can lead to an excessively small learning rate over time, causing the algorithm to converge slowly \cite{zeiler_adadelta_2012}. \\
RMSProp (Root Mean Square Propagation) \cite{tieleman_lecture_2012} addresses the diminishing learning rate problem of Adagrad by replacing the sum of squared gradients with an exponential moving average $\vb_i$: 
\begin{eq}
    \vb_i &= \beta_2 \vb_{i-1} + (1-\beta_2) \gb_i^{\circ 2}  \\
    \lambdab_{i+1} &= \lambdab_i + \eta \gb_i \oslash  \sqrt{\vb_i + \varepsilon}, 
\end{eq}
where $\beta_2$ is a hyperparameter to control to width of the moving average.
This method prevents the learning rate from decreasing too rapidly and enables more balanced optimization progress. \\
Adam (Adaptive Moment Estimation) \cite{kingma_adam_2014} combines RMSProp with a momentum term and additionally accounts for the initialization bias of the moving averages. It computes individual adaptive learning rates using estimates of the first and second moments of the gradients (the mean and the uncentered variance). This dual approach allows it to perform well in practice across a wide range of problems. The update is defined as:
\begin{eq}
    \hat{\mb}_i &= \frac{1}{1-\beta_1^i} \left[\beta_1 \mb_{i-1} + (1-\beta_1) \gb_i \right]  \\
    \hat{\vb}_i &= \frac{1}{1-\beta_2^i} \left[\beta_2 \vb_{i-1} + (1-\beta_2) \gb_i^{\circ 2} \right]  \\
    \lambdab_{i+1} &= \lambdab_i + \eta \hat{\mb}_i \oslash \sqrt{\hat{\vb}_i + \varepsilon}, 
\end{eq}
where $\mb_i$ is the moving average of the gradients, and $\beta_1$ is a hyperparameter to control the exponential decay of the moving average. 
AdaMax \cite{kingma_adam_2014} is a modification to Adam, where instead of dividing the gradients by the moving average of the elementwise $L^2$ norm of the gradients, the $L^{\infty}$ norm is used.
The disadvantage of using moving averages is that in the vicinity of the optimum, the effective learning rate stays almost constant, and the algorithm may not converge.

\subsection{Dynamic Learning Rate Decay (DLRD)}
\begin{figure}[h]
    \centering
    \includegraphics[]{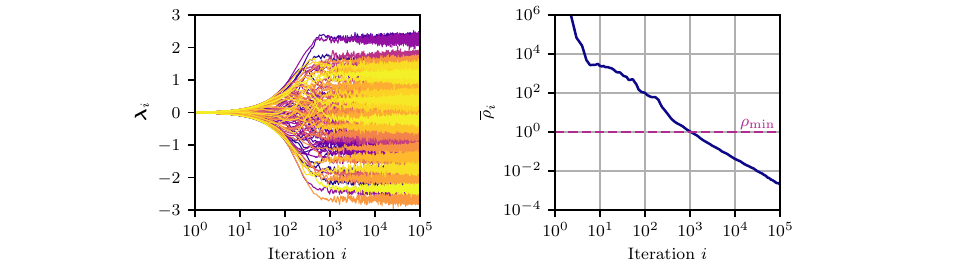}
    \caption{Exemplary course of variational parameters $\lambdab_i$ over iterations $i$ using Adam with static base learning rate and corresponding mean SNR $\overline{\rho}_i$ (see \eqref{eq:snr}).}
    \label{fig:snr}
\end{figure}

Learning rate schedules can help to mitigate this problem by reducing the learning rate over the course of the optimization. Yet, the amount at which the learning rate should be decreased in each iteration is problem-dependent.
To tackle this issue, we propose a method that reduces the step size based on the history of variational parameters. \\
The general idea is to quantify the oscillations relative to the progress of the variational parameters. As a measure for this, we use the signal-to-noise (SNR) ratio related to a linear regression of the variational parameters over the iterations. As we have multiple variational parameters, we compute the SNR $\rho$ (see Appendix \ref{appendix:snr}) for all variational parameters $\lambdab$ in each iteration $i$:
\begin{eq}
    \ab_i &= \sum_{j=0}^i \lambdab_j , \quad \bb_i = \sum_{j=0}^i \lambdab_j^{\circ 2}, \quad \cb_i = \sum_{j=0}^i j \lambdab_j,  \\ 
    \rhob_i &= \left[\frac{i(i+1)(i+2)}{12}\left(\bb_i - \frac{1}{i+1} \ab_i^{\circ 2}\right) \oslash \left(\cb_i - \frac{i}{2}\ab_i\right)^{\circ 2}  -1 \right]^{\circ -1}.
    \label{eq:snr}
\end{eq}
The required additional computational cost and memory increase linearly with the dimension of $\lambdab$ and are comparable to the one required by Adam. Figure \ref{fig:snr} shows an exemplary course of variational parameters over the iterations using Adam with a static base learning rate and the corresponding $\overline{\rho}_i$, which is the average over the SNR vector $\rhob_i$ at iteration $i$. \\
If the average SNR $\overline{\rho}_i$ falls below a defined threshold $\rho_{\mathrm{min}}$, we decrease the base learning rate by multiplying a constant factor $\alpha \in (0, 1)$. When the learning rate is decreased, we reset the linear regression to obtain an SNR solely related to the current learning rate. Further, we prescribe that the number of iterations before the learning rate can be decreased again has to be larger than the previous number of iterations that it took to fall below the threshold in order to increase the stability of the algorithm and prevent excessive reduction of the learning rate. A natural choice for the threshold is $\rho_{\mathrm{min}}=1$, as this indicates that the amount of signal is equal to the noise level. For the decay factor, we propose a default value of $\alpha=0.1$. We found that these values work well in all of our tested examples, such that no further problem-dependent fine-tuning of these hyperparameters should be necessary in most cases. The complete algorithm is presented below.
\begin{algorithm}
\caption{Dynamic Learning Rate Decay ($\textsc{Optimizer}, \lambdab_0, \eta_0, n_{\mathrm{iter}}, \alpha=0.1, \rho_{\mathrm{min}}=1$)}\label{alg:cap}
\begin{algorithmic}
\Require $\ab = \boldsymbol{0}, \bb = \boldsymbol{0}, \cb = \boldsymbol{0}$, $k=-1$, $k_{\mathrm{min}} = 2$
\For{$i=0 \; \textbf{to} \; n_{\mathrm{iter}}$}
\State $k = k + 1$
\State $\ab = \ab + \lambdab_i$ 
\State $\bb = \bb + \lambdab_i^{\circ 2}$
\State $\cb = \cb + k \lambdab_i$
\If{$k \geq k_{\mathrm{min}}$}
    \State $\rhob = \left[\frac{k(k+1)(k+2)}{12}\left(\bb - \frac{1}{k+1} \ab^{\circ 2}\right) \oslash \left(\cb - \frac{k}{2}\ab\right)^{\circ 2}  -1 \right]^{\circ -1}$
    \If{$\overline{\rho} < \rho_{\mathrm{min}}$}
        \State $\eta_i = \alpha \eta_i $
        \State $k_{\mathrm{min}} = k$
        \State $k=-1$
        \State $\ab = \boldsymbol{0}$ 
        \State $\bb = \boldsymbol{0}$
        \State $\cb = \boldsymbol{0}$
    \EndIf
\EndIf
\State $\lambdab_{i+1} = \; $\textsc{Optimizer.step}($i$, $\eta_i$, $\lambdab_i$)
\State $\eta_{i+1} = \eta_i$
\EndFor
\end{algorithmic}
\end{algorithm}

\section{Numerical Examples}
\label{sec:examples}

In the following, we test the performance of our method based on three numerical examples. In the first example, we investigate a synthetic problem with an analytical solution for the optimal variational distribution. Subsequently, we examine Bayesian Logistic Regression on the breast cancer dataset \cite{wolberg_breast_1995, nishihara_parallel_2014}. Finally, we perform Bayesian calibration for a finite element model. In all tested examples, we use SVI with the reparameterization trick, and we use the default values $\alpha=0.1$ and $\rho_{\mathrm{min}}=1$ for the DLRD algorithm.

\subsection{Synthetic Test Case with Generic Joint Distribution}
\begin{figure}[b!]
    \centering
    \includegraphics[]{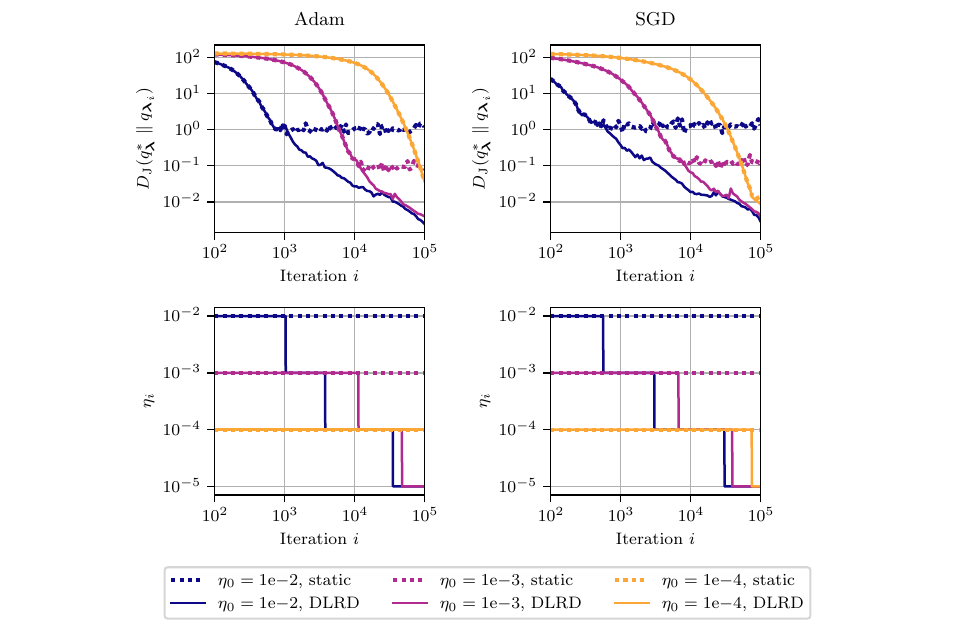}
    \caption{Accuracy of the variational distribution over the number of iterations for the synthetic test case (see \eqref{eqn3:synthetic_joint}) using Adam and SGD optimizer with and without DLRD for varying initial base learning rates $\eta_0$. Accuracy is measured by Jeffreys divergence  between the optimal variational distribution and the variational distribution at the current iteration (see \eqref{eqn:DJ}).}
    \label{fig:quadnormal}
\end{figure}
\begin{figure}[b!]
    \centering
    \includegraphics[]{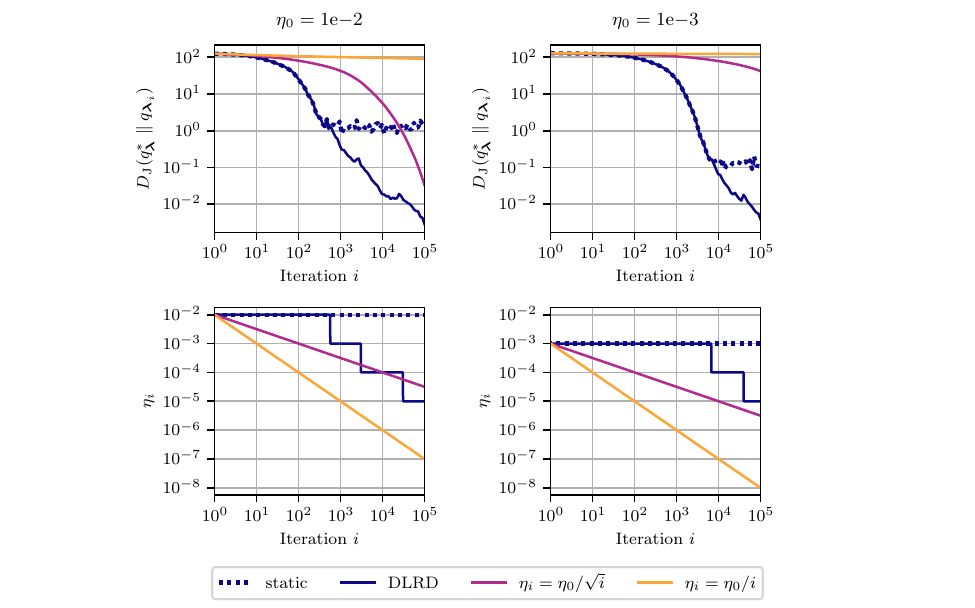}
    \caption{Performance comparison between DLRD and learning rate schedule of the form \eqref{eqn:lr_schedule} with $\zeta=0.5$ (purple) and $\zeta=1$ (yellow) for the synthetic test case (see \eqref{eqn3:synthetic_joint}) using SGD with an initial learning rate of $\eta_0=\num{1e-2}$ and $\eta_0=\num{1e-3}$. Accuracy is measured by Jeffreys divergence between the optimal variational distribution and the variational distribution at the current iteration (see \eqref{eqn:DJ}).}
    \label{fig:quadnormal_loglin}
\end{figure}

\begin{figure}[h!]
    \centering
    \includegraphics[]{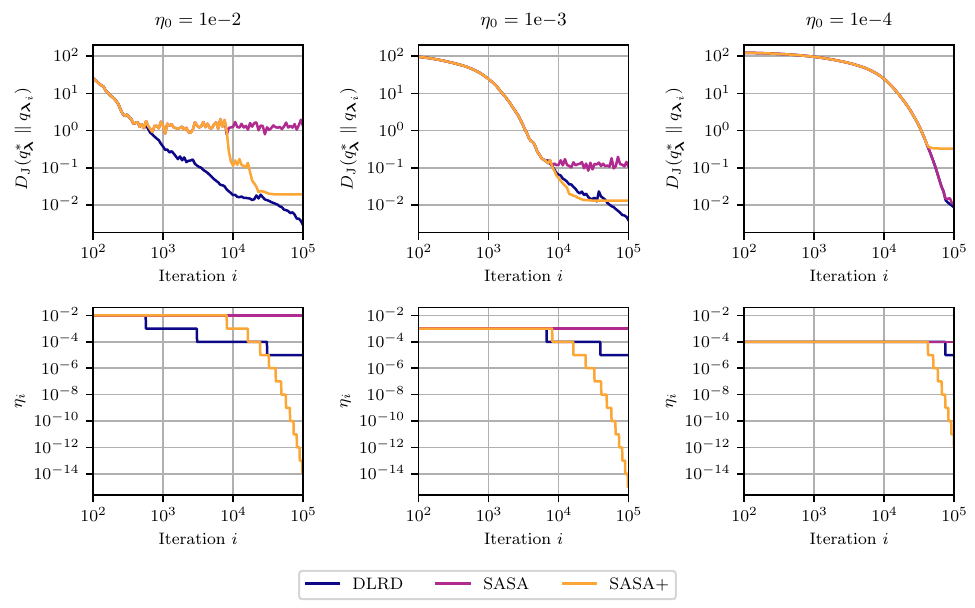}
    \caption{Performance comparison between DLRD, SASA and SASA+ for the synthetic test case (see \eqref{eqn3:synthetic_joint}) using SGD for varying initial learning rates $\eta_0$. Accuracy is measured by Jeffreys divergence between the optimal variational distribution and the variational distribution at the current iteration (see \eqref{eqn:DJ}).}
    \label{fig:quadnormal_sasa}
\end{figure}

\begin{figure}[h]
    \centering
    \includegraphics[]{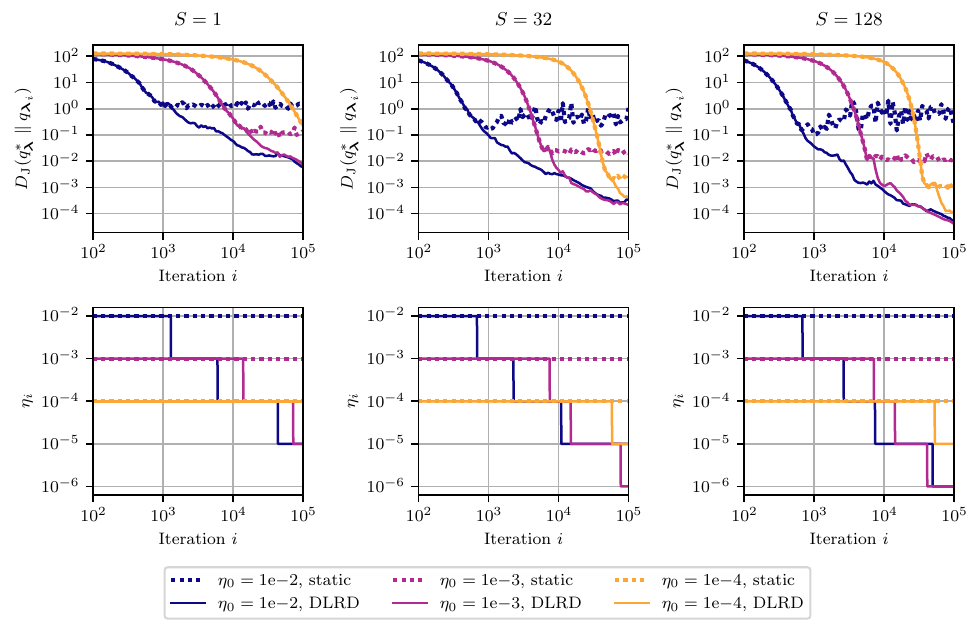}
    \caption{Accuracy of the variational distribution over the number of iterations for the synthetic test case (see \eqref{eqn3:synthetic_joint}) using Adam with and without DLRD for varying initial base learning rates $\eta_0$ and varying batch sizes $S$. Accuracy is measured by Jeffreys divergence between the optimal variational distribution and the variational distribution at the current iteration (see \eqref{eqn:DJ}).}
    \label{fig:quadnormal_sample_size}
\end{figure}

 
In this example, we consider a synthetic test case, where we define the joint distribution $p(\zb, \xb)$ as a product of an unnormalized normal and a generalized normal distribution:
\begin{eq}
    p(\zb, \xb) = \exp \left( -\frac{1}{2} \norm{((\zb - \mub) \oslash \sigmab)^{\circ 4}}_1 -\frac{1}{2} (\zb-\mub)^{\top} \Lambda (\zb-\mub) \right),
    \label{eqn3:synthetic_joint}
\end{eq}
with scaling factor $\sigmab$, precision matrix $\Lambda$ and dimensionality $\zb \in \mathbb{R}^{100}$.
The variational distribution is parameterized as a normal distribution with diagonal covariance. The variational distribution that minimizes the reverse KL divergence in \eqref{eqn:kl} can be found analytically as (see Appendix \ref{appendix:optimal_q}):
\begin{eq}
    q_{\lambdab}^*(\zb) = \mathcal{N}\left(\mub, \frac{1}{12}\diag{\sigmab^{\circ 4} \odot \left(-\diag{\Lambda} + \sqrt{\diag{\Lambda}^{\circ 2}  + 24 \sigmab^{\circ -4}}\right)}\right).
\end{eq}
As a measure of accuracy, we use the symmetric Jeffreys divergence between the current variational distribution and the optimal variational distribution $q^*$:
\begin{eq}
    \DJe{q_{\lambdab}^*(\zb)}{q_{\lambdab}(\zb)} = \DKL{q_{\lambdab}^*(\zb)}{q_{\lambdab}(\zb)} + \DKL{q_{\lambdab}(\zb)}{q_{\lambdab}^*(\zb)}.
    \label{eqn:DJ}
\end{eq}
The KL divergence and, consequently, the Jeffreys difference between two normal distributions can be computed in closed form.\\ 
Figure \ref{fig:quadnormal} shows the accuracy in terms of \eqref{eqn:DJ} over the iterations using Adam and stochastic gradient descent (SGD) with a batch size of $S=2$. For both Adam and SGD, it can be seen that for a large static base learning rate $\eta_0=\num{1e-2}$ (blue dotted line), the optimization progresses rapidly in the beginning but eventually starts to oscillate, leading to an unsatisfying precision of the variational distribution. On the other hand, for a lower static base learning rate $\eta_0=\num{1e-4}$, the optimization progresses slowly. With our proposed DLRD method, the oscillation is detected, and consequently, the learning rate is reduced to reach higher accuracy. For all tested learning rates and both optimizers, the accuracy is improved when the base learning rate is dynamically decreased compared to when it is static. \\
In this example, we also compare the performance of our method to predefined learning rate schedules of the form 
\begin{eq}
    \eta_i = \eta_0 / i^{\zeta}, \quad i \in \mathbb{N}.
    \label{eqn:lr_schedule}
\end{eq}
For $\zeta \in (0.5, 1]$ the Robbins Monroe conditions \cite{robbins_stochastic_1951}:
\begin{eq}
    \sum_{i=0}^{\infty} \eta_i &= \infty  \quad \text{and} \quad \sum_{i=0}^{\infty} \eta_i^2 &< \infty,
    \label{eqn:rm_cond}
\end{eq}
are fulfilled. It can be shown that with the step size requirement in \eqref{eqn:rm_cond} and some additional differentiability and continuity assumptions about the gradient, SGD converges to a local minimum \cite{bottou_optimization_2018}. 
Figure \ref{fig:quadnormal_loglin} depicts the accuracy of the variational distributions over the iterations using SGD with an initial learning rate of $\eta_0=\num{1e-2}$ and $\eta_0=\num{1e-3}$. 
For the learning rate schedule in \eqref{eqn:lr_schedule} with $\zeta=1$, the learning rate decreases rapidly, leading to extremely slow progress of the optimization. With $\zeta=0.5$, the optimization performance is improved due to the slower rate at which the learning rate is decreased. Still, for a lower initial learning rate $\eta_0 = \num{1e-3}$, the optimization progress is relatively slow. Generally, the optimal value of $\zeta$ depends on the problem at hand, the starting point of the optimization, and the initial learning rate. In contrast, our approach does not require hand tuning of additional hyperparameters and still outperforms both the static learning rate as well as the learning rate schedule in \eqref{eqn:lr_schedule} for varying values of $\zeta$. \\
In Figure \ref{fig:quadnormal_sasa}, we compare the performance of DLRD with the statistical adaptive stochastic approximation (SASA) \cite{lang_using_2019} method. In essence, the SASA method performs a statistical test to determine whether the dynamics of SGD have reached a stationary distribution. Once stationarity is detected, the learning rate is dropped by a constant factor, similar to our DLRD approach. The SASA+ \cite{zhang_statistical_2020} method is an extension of SASA that reformulates the statistical test such that it can be applied to a larger variety of stochastic approximation methods. For the comparison, we use SGD for the synthetic test case in \eqref{eqn3:synthetic_joint} with varying initial learning rates $\eta_0$ and use the default hyperparameter values for SASA and SASA+. For SASA, a default value for the learning drop factor is missing, and we therefore use the default drop factor from SASA+, which is also identical to our default drop factor of $\alpha=0.1$. The testing interval for SASA is one training epoch, but as we have no training epochs in SVI, we set the testing interval to $1000$ iterations here.
It can be seen that for all considered initial learning rates $\eta_0$, SASA does not detect stationarity and consequently doesn't reduce the learning rate. Therefore, the behavior is identical to a static learning rate for the considered cases. SASA+, on the other hand, reduces the learning rate too rapidly upon the first detection of stationarity, which leads to a stagnation of the optimization progress. Furthermore, the stationarity is detected rather late for $\eta_0 = \num{1e-2}$ but detected too early for $\eta_0 = \num{1e-4}$. SASA and SASA+ have shown to work well for the training of neural networks, but here, in the context of SVI and the considered example, our DLRD approach performs better for all tested initial learning rates $\eta_0$.  
\\
With increasing batch size $S$, the variance of the gradient estimator reduces, and consequently, the oscillations around the optimum should be reduced. In order to demonstrate that our method works well in practice independent of the batch size $S$, we perform SVI using Adam with varying batch sizes and varying base learning rates as shown in Figure \ref{fig:quadnormal_sample_size}. It can be seen that for all tested configurations, the optimization performance is improved when using the DLRD approach, compared to using a static base learning rate. \\

\subsection{Bayesian Logistic Regression on a Breast Cancer Patient Cohort}
\begin{figure}[h]
    \centering
    \includegraphics[]{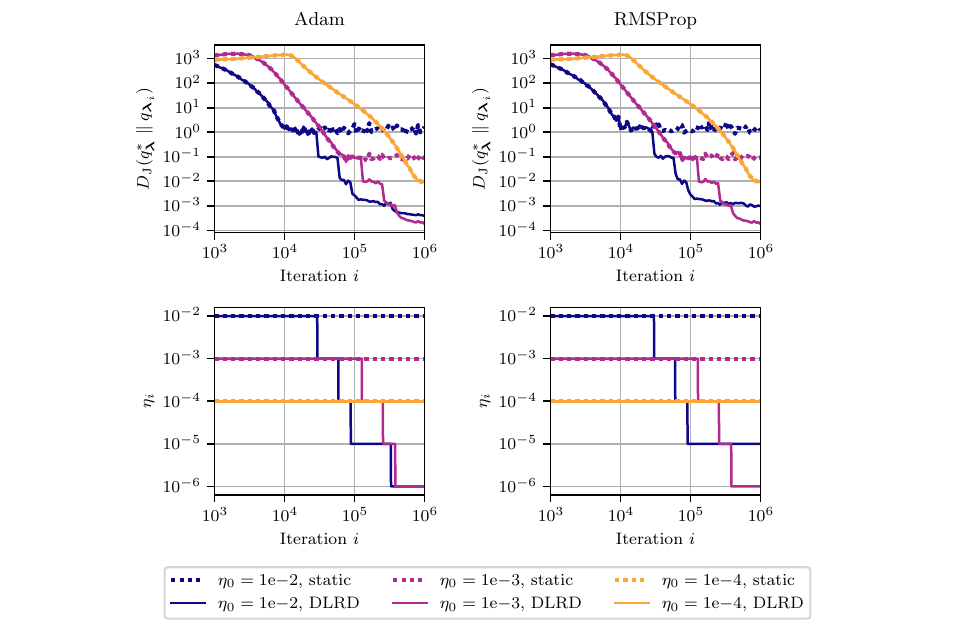}
    \caption{Accuracy of the variational distribution over the number of iterations for the Bayesian logistic regression example using Adam and RMSProp optimizer with and without DLRD for varying initial base learning rates $\eta_0$. Accuracy is measured in terms of the Jeffreys divergence between the approximation of the optimal variational distribution $q^{\mathrm{ref}}_{\lambdab}(\zb)$ and the variational distribution at the current iteration (see \eqref{eqn:DJ}).}
    \label{fig:logistic}
\end{figure}

Similar to \cite{nishihara_parallel_2014, arenz_efficient_2018}, we perform logistic regression on a breast cancer dataset \cite{wolberg_breast_1995} to classify whether the cancer is benign or malignant based on cytological features \cite{wolberg_machine_1994}. The dataset comprises $M=569$ patients with the diagnosis (malignant or benign) as binary targets and 30 real-valued cytological input features. We normalize the input features and add a constant feature to account for the offset, yielding a vector $\thetab_m$ for each patient. The likelihood can be written as:  
\begin{eq}
    p(\xb \mid \zb) &= \prod_{m=1}^{M} \pi_m^{x_m} (1-\pi_m)^{(1-x_m)}, \\ 
    \pi_m &= \frac{\exp \big(\zb^{\top} \thetab_m\big)}{1 + \exp \big(\zb^{\top} \thetab_m \big)}. 
\end{eq}
We can obtain the joint probability density function (pdf) $p(\zb, \xb) = p(\xb \mid \zb) p(\zb)$, where the prior of $\zb$ is  defined as $p(\zb)=\mathcal{N}(\boldsymbol{0}, 100 \, \mathrm{I})$. 
The variational distribution $q_{\lambdab}(\zb)$ is parameterized by the mean and lower Cholesky of the covariance of a normal distribution. Opposed to the previous example, an analytical solution of the optimal variational distribution $q_{\lambdab}^*(\zb)$ is not available here. As a remedy, we compute an approximation $q^{\mathrm{ref}}_{\lambdab}(\zb)$ of the optimal solution $q^*_{\lambdab}(\zb)$ with SVI using Adam with a large batch size of $S=32$ and a large number of iterations ($n_{\mathrm{iter}}=1e7$). We set the initial learning rate to $\eta_0=\num{1e-2}$ and divide the learning rate by a factor of $10$ every $1e6$ iteration. We expect that due to the high initial learning rate, the algorithm does not get stuck in local optima, and due to the high number of iterations in combination with a larger sample size and learning rate reduction, the final iterate should be a very accurate approximation of the optimal solution $q^*_{\lambdab}(\zb)$. Further, we compare the ELBO of the final iterate of multiple optimization runs with different settings to verify that the obtained reference solution is very likely in very close proximity to the optimal solution. \\
Figure \ref{fig:logistic} illustrates the accuracy in terms of the Jeffreys divergence between the approximation of the optimal variational distribution $q^{\mathrm{ref}}_{\lambdab}(\zb)$ and the variational distribution at the current iteration when utilizing Adam and RMSProp with a batch size of $S=8$. Similarly to the previous example, it is visible that with a large static base learning rate $\eta_0=\num{1e-2}$ (blue dotted line), the optimization initially progresses rapidly but subsequently begins to oscillate, resulting in unsatisfactory precision of the variational distribution. Conversely, with a lower static base learning rate $\eta_0=\num{1e-4}$, the optimization advances slowly. With DLRD, the learning rate is reduced when oscillations are detected, which enhances accuracy. For all tested learning rates and both optimizers, accuracy is improved when the base learning rate is dynamically decreased compared to maintaining a static rate. Overall, for the same accuracy, the number of iterations can be reduced drastically when a higher base learning rate with DLRD is utilized, compared to using a smaller and static base learning rate. However, it must be noted that for $\eta_0=\num{1e-3}$ with DLRD, the accuracy is higher than for $\eta_0=\num{1e-2}$ with DLRD after around $\num{3e5}$ iterations. In the case using RMSProp, it can be seen that the learning rate is not reduced after $\num{1e5}$ iterations, as there is still very little progress in the optimization.
Still, the slowdown of the optimization progress for $\eta_0=\num{1e-2}$ is happening at a very high level of accuracy, meaning that it would not be relevant for most practical applications.

\subsection{Bayesian Calibration of a Diffusivity Field}
\begin{figure}[b!]
    \centering
    \includegraphics[]{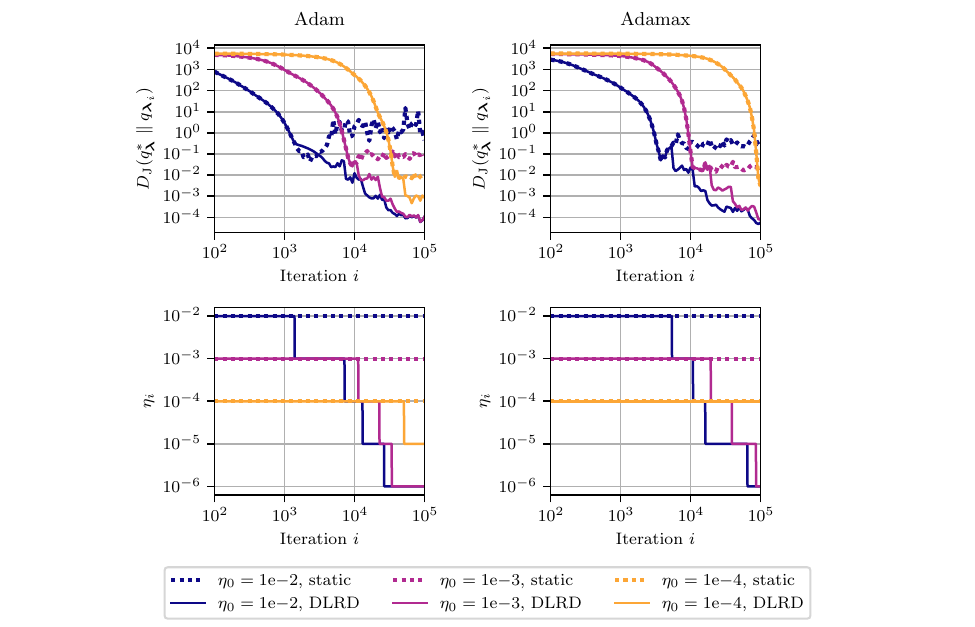}
    \caption{Accuracy of the variational distribution over the number of iterations for the Bayesian calibration example using Adam and AdaMax optimizer with and without DLRD for varying initial base learning rates $\eta_0$. Accuracy is measured in terms of the Jeffreys divergence between the approximation of the optimal variational distribution $q^{\mathrm{ref}}_{\lambdab}(\zb)$ and the variational distribution at the current iteration (see \eqref{eqn:DJ}).}
    \label{fig:diffusion}
\end{figure}
In the third example, we consider a Bayesian calibration problem adapted from \cite{dinkel_solving_2023}, where the goal is to infer a diffusivity field $D$ from observations of a density $u$. The relation between the diffusivity $D(\xib)$ at coordinates $\xib$ and the density $u(\xib)$ is described by the Poisson equation on the unit square domain $\Omega$:
\begin{eq}
   \nabla \cdot (D(\xib) \nabla u(\xib)) &= 10, \quad \xib \in \Omega, \\  
   u(\xib) &= 0,  \phantom{1} \quad \xib \in \partial \Omega. 
   \label{eqn_3_pde_diffusion}
\end{eq}
The partial differential equation (PDE) is solved with the finite element method (FEM) using a mesh of $20 \times 20$ bilinear elements. We use a Gaussian random field $G$ with zero mean and the covariance function:
\begin{eq}
   Cov[G(\xib), G(\xib')] = \exp \left( -\frac{1}{2 l_G^2}\norm{\xib - \xib'}_2^2\right)
   \label{eqn_3_cov_fun}
\end{eq}
to model the diffusivity field as $D(\xib) = \exp\left(G(\xib) \right)$. 
The lengthscale is set to $l_G = 0.2$, and we make use of the Karhunen-Loève expansion \cite{ghanem_stochastic_2003} to discretize the prior of the field utilizing the weighted eigenfunctions of the covariance function $\phi_i$ and coefficients $z_i$:
\begin{eq}
   D(\xib\mid\zb) = \exp\big(G(\xib\mid\zb)\big) = \exp\left(\sum_{i=1}^k z_i \phi_i(\xib)\right).
   \label{eqn_3_kle_diffusion}
\end{eq}
We set the number of expansion terms to $k=29$, such that 99\% of the variance of the field $G$ is preserved. Now we can define the nonlinear mapping $\zb \mapsto \mathcal{M}(\zb)$, which is given by computing the diffusivity $D$ for a given $\zb$ (see \eqref{eqn_3_kle_diffusion}), solving the PDE in \eqref{eqn_3_pde_diffusion} with the FEM, and collecting the values of the density $u$ at defined observation locations $\xib_{\mathrm{obs}}$. \\
The observations $\xb$ are artificially created by drawing a random sample $\zb' \sim p(\zb)$ and adding independent and normally distributed noise with variance $\sigma_n^2$ to the output of $\mathcal{M}(\zb')$. Accordingly, the likelihood takes the form:
\begin{eq}
    p(\xb \mid \zb) &= \mathcal{N}\big(\mathcal{M}(\zb), \sigma_n^2 \mathrm{I}\big). 
\end{eq}
The joint pdf can be obtained as $p(\zb, \xb) = p(\xb \mid \zb) p(\zb)$, where the prior $p(\zb)$ is a standard normal distribution. Here, we use variational inference to approximate the posterior distribution of the diffusivity field coefficients $p(\zb \mid \xb)$.\\
The variational distribution $q_{\lambdab}(\zb)$ is parameterized by the mean and lower Cholesky of the covariance of a normal distribution. With the same argumentation as in the previous example, we expect to obtain an accurate approximation $q^{\mathrm{ref}}_{\lambdab}(\zb)$ of the optimal solution $q^*_{\lambdab}(\zb)$ using Adam with $\eta_0=\num{1e-2}$, $S=32$ samples and dividing the learning rate by a factor of $10$ every $1e5$ iteration until we reach a total number of $1e6$ iterations. \\
Figure \ref{fig:diffusion} illustrates the accuracy in terms of the Jeffreys divergence between the approximation of the optimal variational distribution $q^{\mathrm{ref}}_{\lambdab}(\zb)$ and the variational distribution at the current iteration when utilizing Adam and AdaMax with a batch size of $S=8$. 
As in the previous examples, it is visible that with a large static base learning rate $\eta_0=\num{1e-2}$ (blue dotted line), the optimization begins to oscillate after rapid initial progress. On the other hand, the optimization advances slowly with a lower static base learning rate $\eta_0=\num{1e-4}$. For all tested learning rates and both optimizers, accuracy is improved when the base learning rate is dynamically decreased compared to maintaining a static rate. Moreover, by choosing a higher base learning rate with DLRD, the number of iterations it takes to reach a certain accuracy can be reduced significantly compared to using a smaller static learning rate.

\section{Conclusion}
\label{sec:conclusion}

In this paper, we introduce a novel approach to dynamically reduce the learning rate in Stochastic Variational Inference (SVI) based on the history of variational parameters. Our method addresses the critical issue of oscillating variational parameters near the optimal solution due to a large base learning rate in the stochastic optimization process. By quantifying oscillations relative to the progress of the variational parameters using a signal-to-noise ratio, we enable automatic and adaptive learning rate decay that is both memory-efficient and computationally lightweight. This allows the use of a higher base learning rate, which enables rapid optimization progress without sacrificing accuracy.\\
Through a series of numerical experiments, we demonstrate that our DLRD method effectively prevents optimization stagnation due to oscillations near the optimum. In that sense, our approach significantly mitigates the sensitivity of SVI to the learning rate. Moreover, the results show that our approach not only enhances the performance of SVI when used with standard stochastic gradient descent but also complements existing adaptive learning rate methods such as Adam, AdaMax, and RMSprop. Further, the results indicate that the approach works well independently of the batch size for the gradient estimate. A current limitation of our method is that it can not detect a diverging optimization progress but only oscillations due to high initial learning rates. An algorithm that detects such divergence and resets the optimized parameters to a meaningful state could be complemented with DLRD.  \\  
We believe that our contributions will advance the state-of-the-art in SVI optimization and encourage further research into adaptive learning rate methodologies leveraging the history of variational parameters. As we only use Gaussian families as variational distributions in this work, future research could evaluate the efficacy of our proposed technique for more complex variational distributions, such as mixture models or normalizing flows \cite{rezende_variational_2015}.

\section*{Acknowledgments}
MD and WAW gratefully acknowledge financial support by BREATHE, an ERC-2020-ADG Project, Grant Agreement ID 101021526. GRR and WAW wish to acknowledge support from the German Federal Ministry of Education and Research (project FestBatt2, 03XP0435B).

\printbibliography

\appendix
\section{Derivation of the Signal-to-Noise Ratio}
\label{appendix:snr}
Consider the simple one-dimensional linear regression model with the iteration count $j$ as the independent variable, a variational parameter $\lambda$ as the dependent variable, slope $m$ and offset $t$:
\begin{eq}
    \lambda = m j + t.
\end{eq}
The SNR $\rho$ corresponding to the ordinary least-squares regression model is closely related to the Pearson correlation coefficient $r$ \cite{benesty_importance_2008, rodgers_thirteen_1988}:
\begin{eq}
    \rho &= \left[\frac{1}{r^2}-1\right]^{-1} \\
     &= \left[\frac{\left(\frac{1}{k+1} \sum_{j=0}^k j^2 - \left(\frac{1}{k+1} \sum_{j=0}^k j \right)^2\right) \left(\frac{1}{k+1} \sum_{j=0}^k \lambda_j^2 - \left(\frac{1}{k+1} \sum_{j=0}^k \lambda_j \right)^2\right)}{\left( \frac{1}{k+1} \sum_{j=0}^k j \lambda_j - \left(\frac{1}{k+1} \sum_{j=0}^k j \right) \left(\frac{1}{k+1} \sum_{j=0}^k \lambda_j\right) \right)^2} -1\right]^{-1}\\
    &= \left[\frac{\frac{k (k+2)}{12}  \left(\frac{1}{k+1} \sum_{j=0}^k \lambda_j^2 - \left(\frac{1}{k+1} \sum_{j=0}^k \lambda_j \right)^2\right)}{\left( \frac{1}{k+1} \sum_{j=0}^k j \lambda_j - \frac{k}{2} \left(\frac{1}{k+1} \sum_{j=0}^k \lambda_j\right) \right)^2} -1\right]^{-1}\\
    &= \left[\frac{\frac{k (k+1) (k+2)}{12}  \left(\sum_{j=0}^k \lambda_j^2 - \frac{1}{k+1} \left(\sum_{j=0}^k \lambda_j \right)^2\right)}{\left( \sum_{j=0}^k j \lambda_j - \frac{k}{2} \sum_{j=0}^k \lambda_j \right)^2} -1\right]^{-1}.
\end{eq}

\section{Derivation of the Optimal Variational Distribution}
\label{appendix:optimal_q}
Consider the variational distribution $q_{\lambdab}(\zb) = \mathcal{N}\left(\mb, \diag{\sbo^{\circ 2}}\right)$ and the posterior distribution in \eqref{eqn3:synthetic_joint}. Due to the symmetry and unimodality of $q_{\lambdab}(\zb)$ and $p(\zb, \xb)$, we can conclude the optimal $\mb^* = \mub$.
The ELBO at $\mb = \mub$ can be derived using the identities in \cite{brookes_matrix_2020}:
\begin{eq}
    \mathcal{L} \overset{\mb = \mub}{=} - \frac{3}{2} \norm{(\sbo \oslash \sigmab)^{\circ 4}}_1 - \frac{1}{2} \norm{\sbo^{\circ 2} \odot \diag{\Lambda}}_1 + \frac{1}{2} \norm{ \log \left( \sbo^{\circ 2} \right)}_1 .
\end{eq}
The optimal variance vector $\sbo^{\circ 2}$ can be obtained as:
\begin{eq}
    \nabla_{\boldsymbol{s^{\circ 2}}} \mathcal{L} &\overset{\mb = \mub}{=} - \frac{6}{2} \sbo^{\circ 2} \oslash \sigmab^{\circ 4} - \frac{1}{2} \diag{\Lambda} + \frac{1}{2} \sbo^{\circ -2} = \boldsymbol{0}, \\
    \sbo^{\circ 2} &\overset{\phantom{\mb = \mub}}{=} \frac{1}{12} \sigmab^{\circ 4} \odot \left(-\diag{\Lambda} + \sqrt{\diag{\Lambda}^{\circ 2}  + 24 \sigmab^{\circ -4}}\right),
\end{eq}
where we excluded the negative solution due to positivity constraints. We can see that this is indeed a maximum of the ELBO as the Hessian is negative definite:
\begin{eq}
    \nabla_{\boldsymbol{s^{\circ 2}}} \nabla_{\boldsymbol{s^{\circ 2}}} \mathcal{L} &\overset{\mb = \mub}{=} - \, \diag{3 \sigmab^{\circ -2} + \frac{1}{2} \sbo^{\circ -4}}
\end{eq}

\end{document}